\documentclass[daat]{ipart}

\RequirePackage{hyperref}
\usepackage{xurl}
\usepackage{enumerate}
\usepackage{graphicx}
\usepackage{subfigure}
\usepackage{tikz}
\startlocaldefs
\theoremstyle{plain}
\newtheorem{thm}{Theorem}[section]

\endlocaldefs

\pubyear{2025}
\volume{1}
\issue{1}
\firstpage{1}
\lastpage{4}
\arxiv{0000.0000}

\newtheorem{defn}[thm]{Definition}
\newtheorem{ex}[thm]{Example}
\begin{document}

\begin{frontmatter}
\title[Disentangling Complex Systems: IdopNetwork Meets GLMY Homology Theory]{Disentangling Complex Systems: IdopNetwork Meets GLMY Homology Theory}
%\protect\thanksref{T1}}
%\thankstext{T1}{Footnote to the title with the `thankstext' command.}

\begin{aug}
    \author{\inits{S.}\fnms{Shuang} \snm{Wu}\thanksref{t2}\ead[label=e1]{shuangwu@bimsa.cn}\ead[label=u1,url]{https://www.bimsa.cn/zh-CN/detail/wushuang.html}},
    \address{Beijing Key Laboratory of Topological Statistics and Applications for Complex Systems, Beijing Institute of Mathematical Sciences
     and Applications\\ 
     Beijing 101408, China\\
             \printead{e1}\\
             \printead{u1}}  
    \and
    \author{\inits{M.}\fnms{Mengmeng} \snm{Zhang}\thanksref{t2}\ead[label=e3]{mengmengzhang@bimsa.cn}%
            \ead[label=u2,url]{https://www.bimsa.cn/zh-CN/detail/mengmengzhang.html}}
    \address{Beijing Key Laboratory of Topological Statistics and Applications for Complex Systems, Beijing Institute of Mathematical 
    Sciences and Applications\\
    Beijing 101408, China\\
             \printead{e3}\\
             \printead{u2}}
    \thankstext{t2}{Corresponding author.}
\end{aug}
\received{\sday{6} \smonth{1} \syear{2025}}

\begin{abstract}
The study of complex systems has captured widespread attention in recent years, emphasizing the exploration of interactions and emergent properties among system units. Network analysis based on graph theory has emerged as a powerful approach for analyzing network topology and functions, making them widely adopted in complex systems. IdopNetwork is an advanced statistical physics framework that constructs the interaction within complex systems by integrating large-scale omics data. By combining GLMY theory, the structural characteristics of the network topology can be traced, providing deeper insights into the dynamic evolution of the network. This combination not only offers a novel perspective for dissecting the internal regulation of complex systems from a holistic standpoint but also provides significant support for applied fields such as data science, complex disease, and materials science.
\end{abstract}

%\begin{keyword}[class=AMS]
%\kwd[Primary ]{00K00}
%\kwd{00K01}
%\kwd[; secondary ]{00K02}
%\end{keyword}

%%  Upper case for every keyword

%\tableofcontents
\end{frontmatter}

\section{Introduction}
Growing concern surrounds the interactions and emergent properties on complex systems. Any composition of natural laws can be regarded as a complex system. A complex system is a system composed of many components that may interact with each other~\cite{ladyman2013complex}. Due to the nonlinear, emergent and complex interactions between its components or between a specific system and its environment, it is difficult to model systematically~\cite{bekenstein2003information}. In many cases, the system can be represented in the form of a network, with nodes representing components and links representing their interactions. As a powerful tool for describing the overall characteristics of complex systems, networks have been widely used in research fields such as biology~\cite{auyang1998foundations,kitano2002computational}, medicine~\cite{koithan2012complex}, environmental science~\cite{akjouj2024complex}, and materials science~\cite{chakraborty2024inhibition}.

In modern scientific research, a question of great concern is how to infer real and statistically reliable networks from high-dimensional data and associate them with actual characteristics. With the rise of the big data era, data volume has grown exponentially, leading to the emergence of corresponding data processing methods. Traditional data analysis methods typically follow a reductionist approach, focusing on individual objects to identify single or core components within a complex system~\cite{ladyman2013complex}. However, this approach fails to capture the interactions between different objects. System theory is an anti-reductionist method that emphasizes the interconnections and interdependencies among components, uncovering causal relationships from an overall perspective~\cite{von1968general}. Network is a powerful tool to describe the operation of complex systems and can explain this phenomenon from a system perspective. In the real world, entities may regulate each other in varying directions and intensities, and these relationships can change over time and across different contexts. To address this limitation, Chen et al.~\cite{chen2019omnidirectional} developed the idopNetwork, a general pseudo-temporal physical model that can construct any network with complete information, independent of time-based data.

Network-based methods can systematically characterize the connections between different entities. Currently, numerous graph theory methods have been employed to analyze the structure of complex networks~\cite{newman2003structure}, including degree centrality, betweenness centrality, small-worldness, and network homogeneity~\cite{rosvall2007information,wen2023regulating,xiao2021deciphering}. However, these metrics are often limited by their focus on local properties, typically overlooking the broader topological structure of the network, such as connected components or holes~\cite{aktas2019persistence}. Persistent homology, on the other hand, addresses this limitation by computing the homology groups of a growing object and to track how its homological features evolve along the filtration to capture the global structure or shape of it~\cite{ghrist2008barcodes}. By combining the strengths of network theory and persistent homology, path homology, now known as GLMY homology theory, has emerged as a promising tool for analyzing and studying complex networks, which is specifically applied to weighted directed networks, enabling the capture of quantitative information~\cite{lin2019weighted}.

In this paper, we review the development history of idopNetwork, provide a detailed explanation of its calculation method, and analyze the topological structure of the network results in conjunction with GLMY theory. Currently, the integration of idopNetwork and GLMY theory as an innovative research method has had a significant impact across various fields, including biology, medicine, forestry, and drug design. Finally, we discuss the challenges faced by the model and its future development prospects, with the aim of promoting its broader application in diverse scenarios.

\section{A computational framework for idopNetwork}
Network science is an interdisciplinary field that integrates computational science with various disciplines, playing a crucial role in strengthening the understanding of regulatory relationships within complex systems. Despite its nearly 20-year history, network science remains in its early stages. In recent years, driven by multiple factors, the field has undergone rapid transformations and is facing new computational challenges, including increasing data complexity, the emergence of multi-level data types, and the exponential growth of data volume. These trends highlight the need for continuous evolution in network science research.

Reconstructing multi-level regulatory networks and leveraging them to uncover underlying principles, system behaviors, and functions is a key objective of current research. In biology, various networks such as gene interaction networks, microbial interaction networks, and metabolic networks, exhibiting common structural characteristics: the network is divided into communities or modules, but the connections between modules are sparser than the connections between the underlying nodes~\cite{newman2006modularity}. Inspired by this idea, Wu et al.~\cite{wu2021recovering} proposed a framework for multi-level gene regulatory networks based on the theory of developmental modules. This approach reconstructs interactions between different modules and refines the internal node structure within each module until the number of nodes stabilizes. By applying this model, a high-dimensional, sparse, and module-stable network can be effectively reconstructed~\cite{wu2021recovering}.

In practical modeling, most approaches rely on dynamic datasets with vast amounts of information. However, obtaining high-density and high-precision data entails significant time and labor costs. In biology, network models commonly use time-series data to infer network structures. Yet, existing omics sequencing data often fall short of the requirements for accurate modeling. Considering these constraints in current network research, Chen et al.~\cite{chen2019omnidirectional} developed a statistical framework, idopNetwork, designed to infer an informative, dynamic, omnidirectional and personalized network~\cite{chen2019omnidirectional,wu2021recovering}. Compared to correlation networks and Bayesian networks, idopNetwork offers distinct advantages by integrating principles from niche theory, evolutionary game theory and allometric theory. Leveraging these foundations, idopNetwork can extract dynamic insights from static data, enabling its application across a broad spectrum of system studies. The development of this model introduces new perspectives for advancing network science.

IdopNetwork dissects complex systems as numerous interacting entities and introduces a quasi-dynamic ordinary differential equation (qdODE) system to simulate and reconstruct interaction networks. This system preserves the advantages of time-based ordinary differential equations (ODEs). First, it can capture and predict the dynamic changes in network behavior under varying environmental conditions. Second, given the large number of interacting entities, the resulting regulatory network is inherently multi-layered and complex. Finally, the model enables the extraction of personalized network information for each individual sample across diverse data types, facilitating comparative analysis of network structural changes under different conditions. This approach offers a novel perspective for investigating the dynamic behavior of complex systems.

IdopNetwork has the following characteristics
\begin{enumerate}[(1)]
\item  Informative: The network incorporates bidirectional, weighted, and signed interactions, accurately capturing the diverse relationships between entities. These interactions can be promotive, inhibitive, or neutral, with the strength of each interaction quantitatively expressed to reflect the degree of promotion or inhibition.
\item  Dynamic: The network structure dynamically illustrates how entities within the system respond to environmental changes and signals, adjusting their structure and function accordingly. In gene networks, these changes can be observed by comparing different time points under the same treatment or different treatments at the same time point.
\item Omnidirectional: When the regulatory network involves many entities, it is necessary to establish a multidimensional set of ordinary differential equations to describe the interactions of all entities in the network. The implementation of variable selection can detect the most important nodes, thereby constructing a sparse and stable network.
\item Personalized: Current methods can only perform network inference from high-density time data, and the most prominent advantage of this model is that it can integrate static expression data into a fully informed network. By using the parameters inferred from the overall network, a personalized network can be constructed for each individuals.
\end{enumerate}
The process of idopNetwork is shown as Figure 1.
\begin{figure}[htbp]
    \centering
\includegraphics[width=1\linewidth]{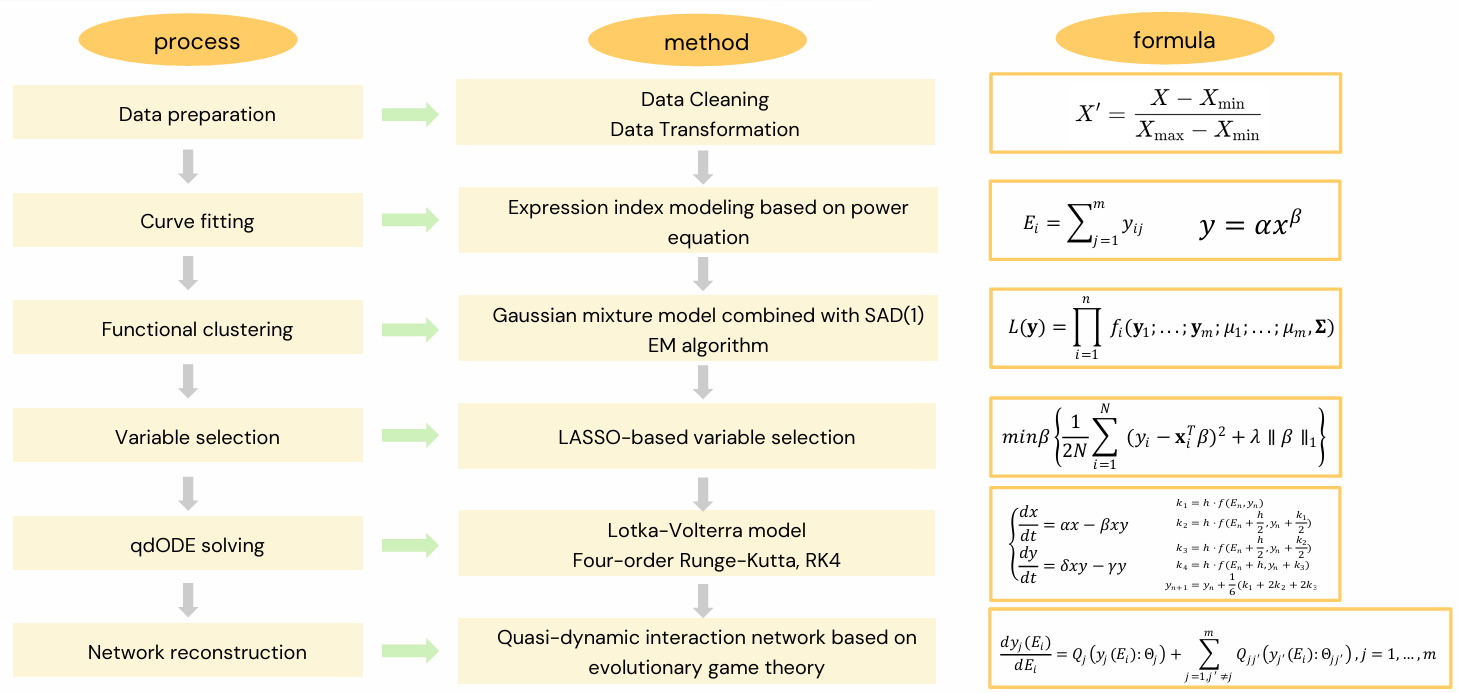}
    \caption{Process of idopNetwork}
    \label{fig:processofidop}
\end{figure}

\subsection{Data structure for IdopNetwork}
Table \ref{Data Structure} presents the input data format required by idopNetwork, which supports both continuous and discrete numerical data. Each column represents a different sample, with a sample size of $n$, where samples may correspond to distinct individuals or repeated measurements from the same individual at different times or locations. Each row represents a different variable, with a total of $m$ variables. In biological contexts, these variables often represent genes, proteins, or metabolites. The sum of each column is defined as the total expression of all variables within a sample, denoted as $E$. Using the power function between individual variables and the total expression, idopNetwork can construct an interaction network that incorporates all variables.
\begin{table}
\caption{Data Structure}\label{Data Structure}
\begin{tabular}{|c|r|r|r|c|}
\hline
&Sample 1& Sample 2& ...&Sample $n$\\
\hline
Variable 1& $y_{11}$&$y_{12}$& &$y_{1n}$\\
\hline
Variable 2& $y_{21}$&$y_{22}$& &$y_{2n}$\\
\hline
\dots&\dots&\dots&&\dots\\
\hline
Variable $m$& $y_{m1}$&$y_{m2}$& &$y_{mn}$\\
\hline
Sum&$E_{1}$&$E_{2}$&$\dots$&$E_{n}$\\
\hline
\end{tabular}
\end{table}

\subsection{Principle for IdopNetwork}
\subsubsection{Niche theory}
In natural ecosystems, various biotic factors (species) interact with abiotic elements (temperature, nutrients, water, oxygen, etc.), creating a complex web of relationships. Viewing organisms, organs, tissues, and even single cells as ecosystems themselves allows us to apply niche theory to explore the internal operational principles of these biological entities. In community ecology, ecosystems are seen as inherently heterogeneous, shaped by interactions among diverse patches. Niche theory helps explain how biotic factors within a shared space interact with those from different spaces, modulating and altering the structure and function of the entire ecosystem~\cite{shea2002community}. Additionally, it clarifies how dynamic changes in niches influence information flow between spaces. In behavioral ecology, species interactions are classified into types such as mutualism, antagonism, aggression, and altruism. These interaction types may also occur microscopically among biological elements like genes, proteins, and metabolites. Applying niche theory enables us to dissect pathogenic mechanisms, modeling them as interaction networks including gene, protein, and metabolic networks within the same spatial context.

During the fitting process, idopNetwork incorporates concepts from niche theory and allometric theory. Specifically, the first step of data analysis in idopNetwork involves selecting appropriate equations and building models. Inspired by niche theory, when describing the full set of genes of an individual, we treat the individual as an ecosystem, with each gene occupying a distinct ecological niche within it, as illustrated in Figure 2. Since each gene represents a portion of the total gene expression of the individual, allometric theory provides an effective way to describe the functional relationship between the expression of a single gene (niche) and the total gene expression of the individual (ecosystem). When data from multiple individuals are available, the total gene expression of each individual can be regarded as an independent variable, analogous to time points commonly used in dynamic data analysis. This leads to a transformation into a pseudo-dynamic framework for further modeling.
\begin{figure}[htbp]
    \centering
\includegraphics[width=0.6\linewidth]{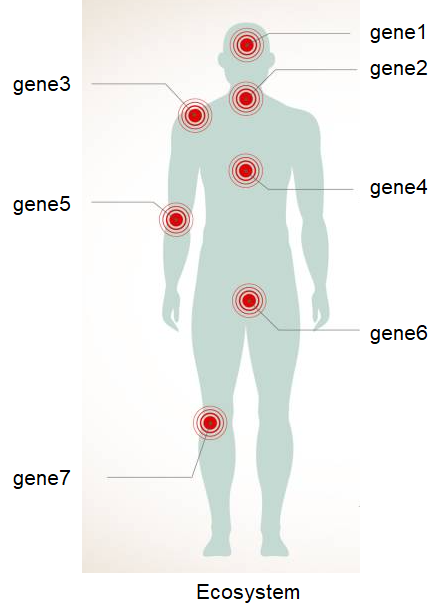}
    \caption{Niche theory for idopNetwork of genes}
    \label{fig:niche}
\end{figure}

\subsubsection{Evolutionary game theory}
The interactions among biomolecules can be understood through the lens of evolutionary game theory. In this framework, biomolecules are seen as individual players in a game, each adjusting its strategy based on the actions of other molecules to optimize its own outcomes. The core concept of game theory, the Nash equilibrium~\cite{nash1950equilibrium}, describes a state where an individual's change in strategy will not yield additional benefits if others maintain their strategies. Extending this, the evolutionarily stable strategy (ESS) adapts the Nash equilibrium concept to scenarios without assuming purely rational behavior, integrating game theory with evolutionary principles.

As molecular interactions evolve over time, evolutionary game theory incorporates dynamic theory to embed ESS within a continuous process, rather than fixing strategy stability to a single moment~\cite{smith1973logic}. Time-based evolutionary game theory has found broad application in fields like complex trait gene analysis, epigenetic regulation of maternal-zygotic transitions, and community ecology in species interactions. This framework offers a comprehensive perspective on molecular interactions and supports a dynamic understanding of biological evolution and ecosystems.

During the network construction phase, idopNetwork incorporates concepts from evolutionary game theory. In nature, phenomena ranging from the population size of a species to the expression level of a gene are not solely determined by the entity itself, but are instead influenced by interactions with other factors. Based on this perspective, we assume that the observed expression level of a gene results from both its independent expression and the effects exerted by other genes. These influences are represented by parameters: a positive parameter indicates a promoting (positive) effect, while a negative parameter indicates an inhibitory (negative) effect. In the network representation, each gene is defined as a node, and the interaction effects between genes are depicted as edges. Red edges represent promotion, whereas blue edges represent inhibition, as illustrated in Figure 3.
\begin{figure}[htbp]
    \centering
\includegraphics[width=1\linewidth]{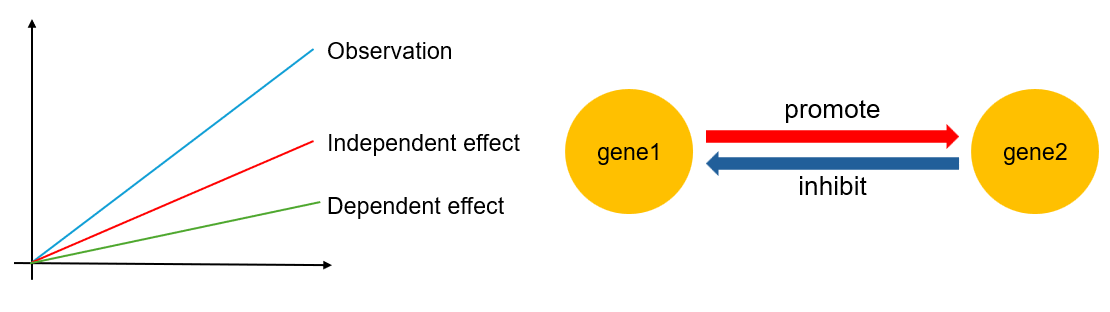}
    \caption{Interactions between genes and their effects}
    \label{fig:interaction}
\end{figure}

\subsubsection{Allometric theory}
Applying evolutionary game theory typically requires dynamic data, yet collecting high-density spatiotemporal omics data presents both economic and ethical challenges. IdopNetwork addresses this issue by incorporating the allometric law as a solution. This law, which describes the relationship between an object’s dimensions and its overall volume as a power function, has been widely used in ecology and physiology to explain biodiversity, biological evolution, and structure-function relationships in organisms~\cite{west1997general}.

Inspired by the environmental index used in agricultural research, this study introduces the concept of a habitat index (HI) for ecosystems. HI quantifies habitat quality by accounting for all biotic and abiotic factors present in a habitat. The relationship between individual biomolecules and HI reflects a part-to-whole dynamic consistent with the allometric law, which can be modeled using a power function. By integrating the allometric law, evolutionary game theory becomes applicable to a broader range of biological contexts, enabling the conversion of static data into dynamic data. This allows researchers to analyze interactions and changes among biomolecules over time, deepening the understanding of biological system complexity and dynamics. Ultimately, this approach provides a novel theoretical foundation for studying biological evolution and ecosystem development.

\subsection{Method of network reconstruction}
\subsubsection{Expression index modeling based on power equation}
Traditional modeling methods typically construct models using time-dependent dynamic data. However, in practical applications, obtaining extensive time-dependent data and samples is often impractical due to the substantial manpower and financial resources required. To address this, our study introduces the allometric growth model in conjunction with niche theory.

In community ecology, the concept of an ecological niche is fundamental, it describes the relationship between organisms and their biological environment across time and space~\cite{shea2002community}. A specific combination of biological factors (such as food resources) at a given time and place defines a point within the niche space. The modern definition of a species' niche encompasses its effects on each point in this space, 
including resource consumption, interference with other organisms' access to resources, and space occupancy~\cite{chesson2000mechanisms}. Ecosystem function and productivity emerge from complex interactions among organisms and between organisms and their environment. Therefore, the total abundance of organisms in an ecosystem, termed the niche index, effectively reflects the cumulative resources and environmental conditions necessary for the survival and growth of these organisms~\cite{peterson2011ecological}, offering a measure of the ecosystem's net quality.

In idopNetwork, the total expression profile from the single-omics data of an individual subject is conceptualized as an ecosystem. Its overall expression level, termed the expression index, is assessed by summing the expression values of all genes, analogous to the concept of a niche index~\cite{ye2019np2}. Assuming that the genomes of $n$ subjects contain $m$ genes, let $y_{ij}$ represent the expression value of gene $j~(j =1, ..., m)$ in subject $i~(i =1, ..., n)$. The expression index  for subject $i$ is then calculated and defined as follows:
\begin{equation}
\label{e1}
   E_{i}=\sum^{m}_{j=1}y_{ij}.
\end{equation}
The allometric model is among the few fundamental laws in biology~\cite{west1997general}. Under natural selection, organisms optimize their metabolic capacity to obtain the energy and resources necessary for survival and reproduction. ~\cite{west1999fourth} demonstrated that this law follows a power-law relationship, grounded in core biophysical and physiological principles. The term ``allometric growth'' was first introduced by~\cite{huxley1936terminology} to study relative growth patterns, and it has since been applied across nearly all measurements of biomass. In living systems, the relationship between parts and the whole is mathematically modeled to adhere to the allometric law, often represented by a power equation~\cite{shingleton2010regulation,west1997general,west1999fourth} as follows:
\begin{equation}
\label{e2}
y=\alpha x^{\beta}
\end{equation}
or
\begin{equation}
\label{e3}
\log y=\log \alpha +\beta \log x.
\end{equation}

The proportional relationship between $y_{ij}$ and $E_i$ can be described by this equation, allowing each gene’s expression level to be represented as a function of the expression index. Here,  $\log \alpha$ denotes the intercept on the $x$-axis, also referred to as the allometric exponent, while $\beta$ serves as the proportionality coefficient. Estimating these parameters enables this study to assess how a subject's gene expression changes in relation to the expression index. This equation provides a framework to model the variation of individual parts as a function of the whole.
\subsubsection{Quasi-dynamic interaction network based on evolutionary game theory}
A defining characteristic of complex systems is the network of interactions among their components and the dynamic and structural shifts these interactions undergo across time, space, and signals. Therefore, a foundational step in deciphering the operational principles of the ecosystem within the human body is to understand how to simulate and estimate the interactions among genes in this ecosystem.

This study posits that gene interaction patterns can be analyzed through the lens of game theory~\cite{von1947theory}, where each gene optimizes its expression based on both its own strategy and the strategies of other genes. A core concept in game theory is Nash equilibrium~\cite{nash1950equilibrium}, which holds that if the strategies of other players remain unchanged, rational players will not benefit from altering their actions. To address the rationality assumption inherent in Nash equilibrium, Maynard Smith and Price (1973) introduced the concept of evolutionary stable strategies, providing a static approach to examine strategy stability within groups with fixed strategies~\cite{bomze1989population}. Evolutionary game theory later refined this concept, offering a dynamic framework to study how strategies evolve over time, unifying standard stability concepts without explicitly defining evolutionary stability~\cite{cressman2014replicator}.

\paragraph{Lotka-Volterra model}
The Lotka-Volterra (LV) model, or predator-prey model, is a foundational mathematical model often used in ecology to describe interactions between predator and prey populations. Proposed in the early 20th century by Italian mathematician Vito Volterra~\cite{volterra1926variazioni} and American mathematician Alfred J. Lotka~\cite{lotka1920analytical}, the LV model employs two differential equations to represent the relationship between predator and prey numbers. It is typically expressed as follows:
\begin{equation}
\label{e4}
\left\{
\begin{aligned}
\frac{dx}{dt}&=\alpha x-\beta x y \\
\frac{dy}{dt}&=\delta x y-\gamma y.
\end{aligned}
\right.
\end{equation}
In this model, $x$ denotes the prey population, $y$ the predator population, and $t$ represents time. The parameters are defined as follows:  $\alpha$ is the prey birth rate, $\beta$ is the predator birth rate, $\gamma$ is the rate at which predators consume prey, and $\delta$ is the predator mortality rate. This model captures predator-prey dynamics, where prey population growth depends on both its own reproductive rate and the predatory pressure, while predator population change is influenced by prey availability and its own growth rate. Such interaction dynamics are common across ecosystems and can be effectively described by the LV model. 
The model can be simplified when studying a single variable within the system:
\begin{equation}
\label{e5}
\frac{dx(t)}{dt}=x(t)(r+ax(t)).
\end{equation}
The LV model is a classic ecological model that uses a simple structure to describe the interactions between organisms in an ecosystem and can serve as a basic framework for understanding the interactions between organisms.
\paragraph{Interacting quasi-dynamic ordinary differential equations}
Various time-based evolutionary game theory models have been developed~\cite{cardelli2017maximal,hart2003uncoupled,hofbauer2009stable}, addressing areas such as the quantitative genetics of complex traits~\cite{fu2018trees}, epigenetic regulation of maternal transitions~\cite{wang2017epigenetic}, and community ecology of interspecific interactions~\cite{zomorrodi2017genome}. However, obtaining time-dependent data remains a frequent challenge in research.

To address this, Chen et al. proposed an interdisciplinary model~\cite{chen2019omnidirectional,sun2021statistical,wu2021recovering} that uses evolutionary game theory to describe interactions between any two entities, where the overall benefit for an entity depends dynamically on its strategic resource acquisition and the external influence exerted by other entities. This process is modeled through a mixed ordinary differential equation (ODE) comprising two components: an entity's intrinsic (independent) expression and the influence (dependent) of other entities. This model serves as a generalized LV framework, where, when multiple entities form a system, an interconnected structure emerges. Here, independent elements (nodes) and interdependencies (edges) form a network, encoding a complete set of directed, signed, and weighted (bDSW) interactions, thus establishing a comprehensive informational network. 

IdopNetwork breaks down the actual expression of a gene into two components: an independent part (representing its intrinsic expression in isolation) and a dependent part (resulting from interactions with other genes). This approach is formulated through a quasi-dynamic ordinary differential equation system (qdODEs). In Equation~(\ref{e1}), by expressing $y$ as a function of $E$, discrete samples can be sequenced, making $y_{ij}$, expressed as $y_{j}(E_i)$, a quasi-dynamic variable that can be modeled in a dynamic framework. Based on the generalized LV model in Equation~(\ref{e5}), the differential equation system can then be redefined as:
\begin{equation}
\label{e6}
\left\{
\begin{aligned}
\frac{dy_{1}(E)}{dE}&=r_1 y_1 (E)+\sum^{d_1}_{j=2} a_{1j}y_{j}(E)\\
\frac{dy_{2}(E)}{dE}&=r_2 y_2 (E)+\sum^{d_2}_{j=1,~j\ne 2} a_{2j}y_{j}(E)\\
  &\quad\vdots\qquad\qquad\vdots\qquad\qquad\vdots\\
\frac{dy_{m}(E)}{dE}&=r_m y_m (E)+\sum^{d_m}_{j=1,~j\ne m} a_{mj}y_{j}(E).
\end{aligned}
\right.
\end{equation}
Simplifying it into the following form:
\begin{equation}
\label{e7}
\begin{aligned}
\frac{dy_{j}(E)}{dE}&=Q_{j}(y_j(E):\Theta_j)+\sum^{m}_{j^{\prime}=1,~j^{\prime}\ne j}Q_{jj^{\prime}}(y_{j^{\prime}}(E):\Theta_{jj^{\prime}}),\\
j&=1,\cdots,m.
\end{aligned}
\end{equation}
In this model, $Q_{j}(y_j(E):\Theta_j)$ represents the independent expression of gene $j$, while $Q_{jj^{\prime}}(y_{j^{\prime}}(E):\Theta_{jj^{\prime}})$ represents the dependent expression of other genes on gene $j$. Here, gene $j$ is influenced by all possible other genes $j^{\prime}~(j^{\prime}=1,\cdots, j-1,j+1,\cdots,m)$. The independent expression of gene $j$ is modeled as a self-regulating function of $y_{j}(E)$ determined by parameter $\Theta_j$, while the dependent expression is expressed as a function of $y_{j^{\prime}}(E)$ influenced by parameter $\Theta_{jj^{\prime}}$.

The qdODEs model is capable of reconstructing a complete interaction network from static data. The network developed in this study is rich in information, dynamic, comprehensive, and personalized, referred to as the idopNetwork~\cite{chen2019omnidirectional,wu2021recovering}. This network has been utilized as a tool to analyze interactions within various complex systems, including microbial community assembly~\cite{cao2022modeling,wang2022vaginal}, genetic structure~\cite{cancers12082086}, and tissue-tissue interactions~\cite{wu2021recovering}.

\paragraph{Quasi-dynamic ordinary differential equation solution}
Equation~(\ref{e7}) represents a system of differential equations where the time derivative is substituted with the derivative of the gene expression index, leading to a quasi-dynamic ordinary differential equation system (qdODEs). The variations in both independent and dependent expressions in Equation~(\ref{e7}) concerning the gene expression index typically lack a straightforward form. To address this, non-parametric methods, such as Legendre orthogonal polynomials (LOP), can be employed to smooth and fit the independent and dependent expressions. The general form of Legendre polynomials is given by:
\begin{equation}
\label{e8}
Q_{r}(\nu)=\sum^{M}_{m=0}(-1)^{m}\frac{(2r-2m)!}{2^r m! (r-m)! (r-2m)!}{\nu}^{r-2m}.
\end{equation}
Since the Legendre polynomials are orthogonal in the interval $[-1,1]$, the gene expression index is first scaled to this interval when calculating actual data.
Equation~(\ref{e7}) after smoothing by Legendre polynomials is relatively complex. This study uses the fourth-order Runge-Kutta method (RK4) to solve the quasi-dynamic ordinary differential equations. RK4 is a commonly used method for numerical solutions of nonlinear ordinary differential equations, especially for first-order ordinary differential equations. In this study, the initial value is assumed to be:
\begin{equation}
\label{e9}
\frac{dy}{dE}=f(E,y),~y(E_0)=y_0.
\end{equation}
The fourth-order Runge-Kutta iteration formula is:
\begin{equation}
\label{e10}
\begin{aligned}
k_1&=h \cdot f(E_n,y_n)\\
k_2&=h \cdot f(E_n+\frac{h}{2}, y_n+\frac{k_1}{2})\\
k_3&=h \cdot f(E_n+\frac{h}{2}, y_n+\frac{k_2}{2})\\
k_4&=h \cdot f(E_n+h, y_n+k_3)\\
y_{n+1}&=y_n+\frac{1}{6}(k_1+2k_2+2k_3+k_4).
\end{aligned}
\end{equation}
In this context, $E_n$ denotes the current time step, represented as the expression index interval, while $y_n$ refers to the gene expression corresponding to this interval. The parameter $h$ indicates the step size, and $k_1$, $k_2$, $k_3$ and $k_4$ represent the slopes calculated at four intermediate steps. The term $y_{n+1}$ signifies the estimated value of gene expression for the subsequent step. After employing the RK4 method to solve the quasi-dynamic ordinary differential equation system, the BFGS algorithm is applied to optimize the function within the framework of the least squares method (LSM) or the maximum likelihood estimate (MLE) to derive the estimated parameter values. 

The least squares method is a method commonly used to fit data. It finds the best relationship between data and model by minimizing the residual sum of squares. The least squares method usually involves calculating the gradient of the objective function and optimizing it using algorithms such as gradient descent to find the optimal parameters. The least squares method is expressed as:
\begin{equation}
\label{e11}
\min_{\theta}\sum^{n}_{i=1}[y_i-f(x_i;\theta)]^2.
\end{equation}
Among them, $\theta$ is the parameter to be solved, and the parameter $\theta$ that minimizes the sum of squared residuals is found by solving a system of equations whose partial derivatives with respect to the parameters are equal to zero. 
Maximum likelihood estimation (MLE) is a widely used parameter estimation method in statistics. The fundamental principle of MLE is to select the parameter values that maximize the likelihood of observing the given data. Let $\mathbf{y_j}=(y_j(E_1),\cdots, y_j(E_n))$ represent the expression of gene $j~ (j =1, ..., m)$ measured across $n$ different samples (subjects). Assuming that $m$ genes interact in a manner based on the quasi-dynamic ordinary differential equations (qdODEs), the likelihood of the gene expression data collected from the $n$ subjects can be expressed as:
\begin{equation}
\label{e12}
L(\mathbf{y})=\prod^{n}_{i=1} f_i(\mathbf{y_1};\cdots;\mathbf{y_m};\mu_1;\cdots;\mu_{m^{\prime}};\mathbf{\Sigma}),
\end{equation}
%%%%%%%% please pay attention to this para, there are some missings
where $f_i(\cdot)$ is the $m$-dimensional longitudinal normal probability density function, the mean vector is $\mu=(\mu_1,\cdots,\mu_m)$, and the covariance matrix is $\mathbf{\Sigma}$. The likelihood values of independent expression and dependent expression of all genes are calculated by RK4, and the values of the independent part are regarded as the nodes of the network, and the values of the dependent part are regarded as the edges of the network, and an interaction network with direction, weight and sign is obtained. This network derived by maximum likelihood estimation is stable in network topology.
\subsubsection{Network sparsity}
Not all genes in a biological system are interconnected, resulting in a sparse interaction network. To ensure this sparsity, two strategies can be employed~\cite{wu2021recovering}.

The first strategy involves reducing the number of interactions between genes. Various statistical methods, including regularized variable selection, can be used to identify a small set of genes that are closely related to a given gene. Through variable selection, a fully interconnected network is transformed into a sparsely interconnected one, thereby simplifying the network's complexity. Based on the variable selection described in Equation~(\ref{e6}), the total expression of $m$ genes in the dependent part of gene $j$ can be reduced to the sum of the expression of $d_j$ genes, where gene $j$ is only associated with $d_j~ (d_j \ll m)$ other genes. This achieves network sparsity by solving the simplified qdODEs system.

The second strategy focuses on decomposing a large network into distinct sub-networks and further breaking down each sub-network into smaller sub-sub-networks. This process is repeated until the number of genes in each sufficiently small sub-network is manageable. While genes from different sub-networks can interact, the degree of interaction is typically lower than that among genes within the same sub-network. Specifically, all genes are clustered into different modules, each containing fewer genes, with stronger connections among genes within the same module than those between genes from different modules. This approach to network decomposition aligns with developmental module theory~\cite{melo2016modularity}, which posits that such structures enhance the stability and robustness of living systems in the face of environmental changes.
In this study, both strategies are employed to reconstruct the sparse network of each module. The inter-module network is determined by summing the expression levels of all genes within each module. By linking the module-module network with its derived gene-gene network, multi-level idopNetworks are reconstructed. The top layer features a coarse-grained network, where different sub-networks connect through variable selection. The bottom layer consists of a fine-grained network made up of individual genes. Positioned between the top and bottom layers is an interaction network reconstructed from sub-networks of varying dimensions.
\paragraph{Variable selection}
In real organisms, it is impossible for all genes to interact with each other, so variable selection and dimensionality reduction are performed on the interaction relationship between genes. LASSO (Least Absolute Shrinkage and Selection Operator) is a commonly used statistical method for model sparsification and feature selection. It penalizes the coefficients of the model by adding $\ell_1$ regularization terms to the loss function~\cite{santosa1986linear}. LASSO constrains the model parameter range by using the $\ell_1$ of the coefficient vector, making some coefficients zero, thereby simplifying the model. It can be expressed as:
\begin{equation}
\label{e13}
\min\beta\{\frac{1}{2N}\sum^{N}_{i=1}(y_i-\mathbf{x}^{T}_{i}\beta)^2+\lambda ||\beta||_1\}.
\end{equation}
%%% still some missing
Where $\beta$ is the parameter vector of the model, $N$ represents the number of samples, $y_i$ and $\mathbf{x_i}$ represent the target value and feature vector of the $i$-th sample, respectively, $\lambda$ represents the regularization parameter, which controls the strength of the regularization term, and $ ||\beta||_1$ represents the $\ell_1$ norm, which is the sum of the absolute values of the parameter vector. For the actual data analysis, computations are performed using the $\mathbf{glmnet}$ function in the R software.
\paragraph{Functional clustering}
In network systems, the complexity of complex systems is formed by dense interactions from underlying components. Modularity theory shows that networks can be naturally divided into communities or modules~\cite{cantini2015detection,newman2006modularity}, and the sparsity between different parts is an important condition for maintaining system stability and system adaptation to environmental changes~\cite{ravasz2002hierarchical,valverde2017evolution}. 

Genes with similar expression patterns may have similar functions. In a network system, genes with similar functions should belong to the same module, and genes in the same module often have closer interactions. This study uses a functional clustering method based on a mixed model to cluster all genes in single-omics data. According to the experimental design in Equation~(\ref{e1}), $m$ genes are grouped into L modules according to their expression patterns with the total gene expression index $(E_1,\cdots, E_n)$. The likelihood function of functional clustering is expressed as:
\begin{equation}
\label{e14}
L(\mathbf{y})=\prod^{m}_{j=1}[\Pi_1 f_1(\mathbf{y_j})+\cdots+\Pi_jf_l(\mathbf{y_j})].
\end{equation}
Among them,  $\Pi$ is the proportion of variables belonging to module $l~(l=1,\cdots,L)$, $f_{l}(\mathbf{y}_j)$ is the $n$-dimensional normal distribution probability density function of variable $j$ on the total expression index, the mean vector is $\mu_l=(\mu_l(E_1),\cdots,\mu_l(E_n))$, and the covariance matrix is $\Sigma$. Using Equation~(\ref{e2}) to model the mean vector, it can be expressed as:
\begin{equation}
\label{e15}
\mu_l=(\mu_l(E_1),\cdots,\mu_l(E_n))
=(\alpha_l E^{\beta_l}_{1},\cdots,\alpha_{l}E^{\beta_l}_{n}).
\end{equation}
In this context, $\alpha_l$ and $\beta_l$ represent the exponential coefficients of the sum of the gene expression levels in module \( l \), collectively defining the expression pattern of all genes within that module. The covariance matrix is modeled using the first-order SAD(1) model. The module proportions, along with the parameters for the mean vector and covariance matrix, can be estimated through maximum likelihood estimation using the Expectation-Maximization (EM) algorithm. 

The initial parameters are obtained by the kmeans algorithm, which is a commonly used clustering method that classifies by calculating the distance between each data point and its nearest mean cluster.

In step E, the posterior probability of each variable $j$ belonging to a specific module $l$ is calculated by the formula:
\begin{equation}
\label{e16}
\Pi_{ij}=\frac{\Pi_l f_l(\mathbf{y}_j)}{\Pi_1 f_1(\mathbf{y}_j)+\cdots+\Pi_L f_L(\mathbf{y}_j)}.
\end{equation}
In the M step, the proportion of genes in module $l$ to all genes is estimated:
\begin{equation}
\label{e17}
\Pi_l=\frac{1}{m}\sum^{m}_{j=1}\Pi_{bj}.
\end{equation}
In the M step, the simplex algorithm is used to estimate the parameters in the mean vector and covariance matrix. Repeat the E and M steps until a stable estimate is obtained. Estimate the posterior probability of each gene according to Equation~(\ref{e15}), and assign the gene to the module with the highest probability among the L different modules. 
The optimal number of modules L can be determined by the information criterion. $\rm{AIC}$ (Akaike Information Criterion) and $\rm{BIC}$ (Bayesian Information Criterion) are both commonly used criteria for model selection, which are used to select the optimal model given a data set.
\begin{align}
{\rm AIC}=2k-2\ln(\hat{L})\label{e18};\\
{\rm BIC}=k\ln(n)-2\ln(\hat{L})\label{e19}.
\end{align}
In the AIC model, $k$ represents the number of parameters in the model, and $\hat{L}$ is the maximum likelihood estimate of the model; in the BIC model, $k$ represents the number of parameters in the model, $n$ represents the sample size, and $\hat{L}$ is the maximum likelihood estimate of the model. Compared with the AIC model, the BIC model is more complex. For both criteria, the smaller the AIC and BIC, the better the model selection.

\section{GLMY Homology Theory --- Asymmetry Representations and Fingerprints}
The high-order interactions among units are ubiquitous in both the microscopic and macroscopic worlds. To capture the pairwise and group interactions as they naturally play, simplicial complexes are the simplest choice~\cite{battiston2021physics,battiston2020networks,grilli2017higher}. They are also a classical and core concept in algebraic topology~\cite{hatcher2005algebraic}, with their homology groups serving as effective and computable fingerprints to capture the global structure or shape of complex systems. This gave rise to the fields of Topological Data Analysis (TDA) ~\cite{carlsson2009topology,edelsbrunner2002topological,verri1993use} and Applied Topology~\cite{ghrist2008barcodes,ghrist2014elementary} about two decades ago, which focus on computing the persistent homology groups of evolving objects and tracking how their homological features evolve along filtration~\cite{battiston2020networks,carlsson2009topology,ghrist2014elementary}, offering qualitative, quantitative, and stable features (fingerprints) amidst random noise. However, when it comes to asymmetric or directed interactions, simplicial complexes seem powerless. The digraph and its GLMY homology theory, introduced by Grigor'yan, Lin, Muranov, and Yau in 2012~\cite{grigor2012homologies}, overcome this limitation and have demonstrated their strength and effectiveness in various fields. 
\subsection{GLMY homology theory} 
For asymmetric or unbalanced interactions, a directed graph (digraph) is an effective representation that captures the communication between nodes~\cite{chowdhury2018persistent}. From this perspective, digraphs is everywhere. In this paper, a \emph{digraph} $G$ consists of a set of vertices $V_G$ and a set of arrows $A_G$, where $A_G$ is a subset of $V_G\times V_G$ excluding the diagonal elements. In other words, a digraph does not have multiple arrows between the same pair of vertices or self-loops. In 2012, the path complex and path homology group~\cite{grigor2012homologies}, recalled as GLMY homology group, on digraphs was introduced by Grogor'yan  et al., which helps us analyze and uncover the essential structural information as dimensional fingerprint. In recent years, path homology theory on digraphs has led to the emergence and expansion of digraph topology theory~\cite{di2024path,grigor2012homologies,grigor2014homotopy,grigor2018fundamental,grigor2018path,grigor2015cohomology,grigor2017homologies,grigor2020path} and hypergraph topology theory~\cite{bressan2016embedded} and has inspired numerous advancements and breakthroughs in fields such as biology~\cite{che2024idopnetwork}, chemistry~\cite{chen2023path}, complex networks~\cite{chowdhury2019path,chowdhury2022path}, and complex diseases~\cite{wu2023metabolomic}.

Besides, the path complex is a mathematical generalization of the simplicial complex, which implies that it can adapt more flexibly to a wider range of practical problems without being constrained by the same limitations of simplicial complex.  

Let \( V \) be a nonempty finite set. An \emph{elementary $n$-path}~\cite{grigor2012homologies} on $V$ is a sequence of vertices $v_0v_1\cdots v_n$, $v_i \in V$, $0\leq i\leq n$. A \emph{path complex} on \( V \) is a sequence of sets $\{P_n\}_{n\geq 0}$ of elementary paths on \( V \) such that for each $v_0 v_1 \cdots v_n \in P_n$, $v_0 v_1 \cdots v_{n-1} \in P_{n-1} \text{ and } v_1 v_2 \cdots v_{n} \in P_{n-1}$. Undoubtedly, a simplicial complex is always a path complex. 
Let $G$ be a digraph.  The \emph{elementary path complex} $\Lambda(V_G)$ of $G$ is a sequence of vector spaces  $\{\Lambda_n(V_G)\}_{n\geq 0}$ and a sequence of boundary operators $\{\widetilde{\partial}_n:\Lambda_n(V_G)\rightarrow \Lambda_{n-1}(V_G)\}_{n\geq 0}$, where $\Lambda_n(V_G)$ is the vector space generated by the elements $e_{v_0v_1\cdots v_n}$ indexed by all elementary $n$-paths $v_0v_1\cdots v_n$ in $V_G$ and   $$\widetilde{\partial}_n : \Lambda_n(V_G)\rightarrow\Lambda_{n-1}(V_G),$$ $$  ae_{v_0v_1v_2\cdots v_{n}}\mapsto \sum\limits_{i=0}^{n}(-1)^i ae_{v_0v_1v_2\cdots v_{i-1}\widehat{v_i}v_{i+1}\cdots v_n}$$ for $n\geq 1$, where \( e_{v_0v_1v_2\cdots v_{i-1}\widehat{v_i}v_{i+1}\cdots v_n} \) denotes the \((n-1)\)-dimensional element in \(\Lambda_{n-1}(G)\) indexed by the \((n-1)\)-path obtained by deleting the vertex \( v_i \), 
 and $$\widetilde{\partial}_0 : \Lambda_0(V_G)\rightarrow 0,$$ that is to say, $\Lambda_{-1}(V_G) = 0$ and $\widetilde{\partial}_0$ is a zero homomorphism. 

An elementary $n$-path $v_0v_1\cdots v_n$ is called \emph{non-regular} if there exists $0\leq i\leq n-1$ such that $v_i = v_{i+1}$. On the other hand, the path $v_0v_1\cdots v_n$ is called \emph{regular}. Let $I_n(V_G)$ denote the vector space generated by the elements indexed by all non-regular $n$-paths and $R_n(V_G)$ denote the vector space generated by the elements indexed by all regular $n$-paths. It is easily to check that the boundary operator $$\widetilde{\partial}_n: I_n(V_G)\rightarrow I_{n-1}(V_G)$$ is well-defined on $I_n(V_G)$ as a sub-vector space of $\Lambda_n(V_G)$.  Inheriting the property $\widetilde{\partial}_{n-1} \circ \widetilde{\partial}_n = 0$, the boundary operator $$\partial_n:\Lambda_n(V_G)/I_n(V_G) \rightarrow \Lambda_{n-1}(V_G)/I_{n-1}(V_G)$$ induced by $$\widetilde{\partial}_n: \Lambda_n(V_G) \rightarrow \Lambda_{n-1}(V_G)$$ is well-defined. Therefore, ($\{\Lambda_{n}(V_G)/I_{n}(V_G)\}_{n\geq 0},\{\partial_{n}\}_{n\geq0}$) forms a chain complex. On the other hand, for the sub-vector space $R_{n}(V_G)$ of $\Lambda_n(V_G)$,  $$\widetilde{\partial}_n:R_{n}(V_G)\rightarrow R_{n-1}(V_G)$$ is not well-defined. For example, path $e_{010}$ is a regular 2-path and its boundary  $\widetilde{\partial}_2(e_{010})= e_{10} - e_{00}+e_{01}$. 
Clearly, $e_{00}$ is not an element of $R_1(V_G)$. Fortunately, $R_n(V_G)  \cong\Lambda_n(V_G)/I_n(V_G)$. By using this isomorphism, $(\{R_{n}(V_G)\}_{n\geq 0}, \{\partial_{n}\}_{n\geq 0})$ can form a chain complex under the quotient boundary operator 
\begin{footnotesize}$$\partial_n\colon R_n(V_G)\to \Lambda_n(V_G)/I_n(V_G)\to \Lambda_{n-1}(V_G)/I_{n-1}(V_G) \to R_{n-1}(V_G).$$
\end{footnotesize} From now on, we will refer to $R(V_G)$ as $(\{R_{n}(V_G)\}_{n\geq 0}, \{\partial_n\}_{n\geq0})$.  A simple example about $R_n(V_G)$ and $\partial_n$ is given as follows.
\begin{ex}
Let $G$ be the line digraph shown as follows.
\begin{figure}[htbp]
    \centering
\includegraphics[width=0.3\linewidth]{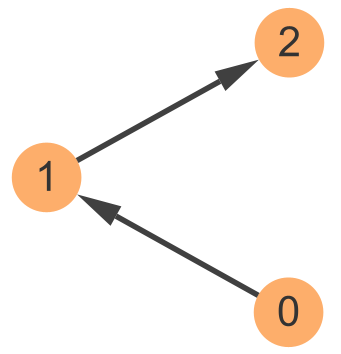}
    \caption{Line digraph $G$}
    \label{fig:digraph}
\end{figure}

$$R_0(V_G) = \langle e_0,\text{ }e_1\text{ },e_2\rangle ;$$
$$R_1(V_G)=\langle e_{01},\text{ }e_{02}, \text{ }e_{10}, \text{ }e_{12},\text{ }e_{20},\text{ }e_{21}\rangle ;$$ $$R_2(V_G) = \langle e_{010},\text{ }e_{012},\text{ }e_{020}, \text{ }e_{021}, \text{ }e_{101},\text{ }e_{102},\text{ }e_{120},$$ $$\text{ }e_{121},\text{ }e_{201},
\text{ }e_{202}, \text{ }e_{210},\text{ }e_{212}\rangle. $$ 
By definition, $\widetilde{\partial}_n(e_{010}) = e_{10}-e_{00}+e_{01}$. But since 1-path $e_{00} \in I_1(V_G)$, then $e_{00}$ is identified with zero element in the quotient space $\Lambda_1(V_G)/I_1(V_G)$. Hence,  $\partial_n(e_{010}) = e_{10}+e_{01}$.
\end{ex} Obviously, the regular path complex $R(V_G)$ disregards the the arrowed information of digraph $G$, and so does elementary path complex $\Lambda(V_G)$. Taking the arrowed information of the digraph into account, the concept of an allowed path was introduced by Grigor'yan, Lin, Muranov, and Yau~\cite{grigor2012homologies}. An $n$-path $v_0v_1\cdots v_n$ of $G$ is called \emph{allowed} if $v_iv_{i+1}\in A_G$, $0\leq i\leq n-1$. Denote  $A_n(G)$ the sub-vector space of $R_n(V_G)$ spanned by the elements indexed by all allowed $n$-paths. Nevertheless, it does not necessarily hold that $\partial_n(A_n(G))\subseteq A_{n-1}(G)$. For example, the boundary $\partial_2(e_{012})$ of allowed $2$-path $e_{012}$ of $G$ in Figure 4 is $e_{12}-e_{02}+e_{01}$, in which $e_{02}$ is not an element of $A_1(G)$. Due to this fact, they consider the following subspace
\[
\Omega_n(G) := A_n(G) \cap \partial_n^{-1}(A_{n-1}(G))
\]
of \( A_n(G) \), where \( \partial_n^{-1}(A_{n-1}(G)) \) denotes the preimage of \( A_{n-1}(G) \) under the homomorphism \( \partial_n \), and the \emph{$\partial$-invariant chain complex} $\Omega(G)=(\{\Omega_{n}(G)\}_{n\geq 0}, \{\partial_{n}(G)\}_{n\geq0})$. 
In particular, \(\Omega_0(G)\) and \(\Omega_1(G)\) are the vector spaces generated by the sets \(V_G\) and \(A_G\), respectively.
  
Further, based on this $\partial$-invariant chain complex, the path homology group is given.
\begin{defn}\label{pathhomo}Let $G$ be a digraph. The \emph{$n$-dimensional path homology group} $H_n(G)$ is defined by $$H_n(G) = H_n(\Omega(G)) = \frac{\mathrm{ker} (\partial_n|_{\Omega_n(G)})}{\mathrm{Im}(\partial_{n+1}|_{\Omega_{n+1}(G)})}.$$ 
\end{defn}
From the definition of $\partial$-invariant chain complex, the path homology groups capture the global structure of $G$, rather than local vertex information. And combining with Theorem 3.16 in~\cite{grigor2018path2}, the path homology group is a homotopy invariant that reflects the structural features of digraph~$G$. 
The reader is referred to the works \cite{grigor2012homologies,grigor2014homotopy,grigor2018fundamental,grigor2018path,grigor2015cohomology,grigor2017homologies,grigor2020path} of Grigor'yan et al. Further, to precisely map the real world, the weighted path homology group has been studied in~\cite{lin2019weighted}.
% Given that a simplicial complex is a special case of a path complex, the path homology group reveals certain "hole" structures on the directed graph.
\begin{ex}
Let $G$ be the following triangle digraph. Then $$A_0(G) = \langle e_1,e_2,e_3\rangle;\quad \quad \quad A_1(G) = \langle e_{12},e_{23},e_{31}\rangle;  \quad \quad \quad $$ $$A_2(G) = \langle e_{123},e_{231},e_{312}\rangle.\quad \quad \quad $$
\begin{figure}[htbp]
    \centering
\includegraphics[width=0.3\linewidth]{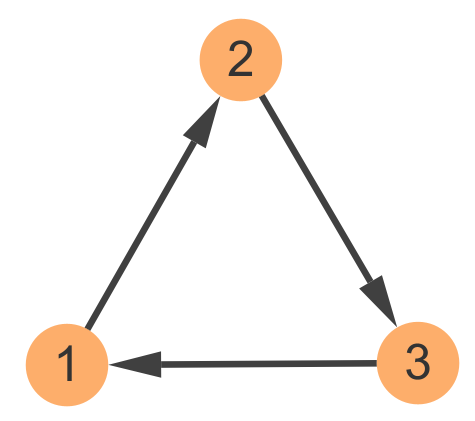}
    \caption{Triangle $G$}
    \label{squ}
\end{figure}

By computation, $$\Omega_0(G) = \langle e_1,e_2,e_3\rangle;\quad \Omega_1(G) = \langle e_{12},e_{23},e_{31}\rangle;\quad \Omega_2(G) = 0.$$ Then the path homology groups with coefficient $\mathbb{Z}/2$ are 
$$H_0(G) = \frac{\mathrm{ker} 
 \partial_0}{\mathrm{Im}\partial_1} = \frac{\Omega_0(G)}{\mathbb{Z}/2\langle e_2-e_1,e_3-e_2\rangle} = \mathbb{Z}/2;$$
 $$H_1(G) = \frac{\mathrm{ker} 
 \partial_1}{\mathrm{Im}\partial_2} = \frac{\mathbb{Z}/2\langle e_{12}+e_{23}-e_{31}\rangle}{0} = \mathbb{Z}/2;$$
 $$H_2(G) = 0.$$
\end{ex}
In recent years, inspired by path homology, the embedded homology of hypergraphs~\cite{bressan2016embedded} and super-hypergraphs~\cite{grbic2022aspects}, introduced by Bressan et al. and Grbi\'c et al., have been developed as a generalization of path (GLMY) homology. These homology theories are applicable to both point cloud data and graph data with the attribute of being adaptable to more wider range of applications. A \emph{hypergraph} $\mathcal{H}$~\cite{berge1985graphs} on a vertex set $V$ is a set of subsets of $V$ without any limitation. A \emph{super-hypergraph}~\cite{grbic2022aspects} is a generalization of hypergraph that allows multi-hyperedges.   

\subsection{Persistent GLMY homology for digraphs}
Persistent homology has already demonstrated its strong and wide power~\cite{battiston2021physics,lawson2019persistent}  to detect and
 analysis intrinsic global geometric and topological information, which
made topology as one of the most commonly used mathematical tools in Data Science~\citep{edelsbrunner2014short,edelsbrunner2013persistent}.  
Inspired by this, and to analyze patterns in directed networks, Chowdhury and M\'emoli~\cite{chowdhury2018persistent} introduced persistent path homology, bridging classical persistence methods and more sufficient directed structures. They also provided an algorithm for persistent path homology and a proof of its robustness in their paper. This facilitates the study of instinct topological features in graph data, which provides a new insight to researchers to dig into data from complex systems. Briefly speaking, a \emph{filtration} $f:\mathcal{P}\rightarrow \mathbb{R}$ on path complex is a function such that $f(\sigma)\leq f(\tau)$ for any paths $\sigma< \tau $, where \( < \) denotes that \( \sigma \) is a sub-path of \( \tau \). 
Then there is a filtration of path complexes $$\emptyset = \mathcal{P}_0 \subseteq \mathcal{P}_1 \subseteq \cdots \subseteq  \mathcal{P}_m \subseteq \cdots,$$ where $\mathcal{P}_i= f^{-1}((-\infty,i])$. In practice, it is typically assumed that after finitely many steps, \( \mathcal{P}_m = \mathcal{P} \). That is,  
\[
\emptyset = \mathcal{P}_0 \subseteq \mathcal{P}_1 \subseteq \cdots \subseteq \mathcal{P}_m = \mathcal{P}.
\]
 Examples include arrow filtration and distance filtration. 
\begin{ex}[Arrow filtration]
    Let \( G \) be a weighted digraph, that is, a digraph equipped with a function $w: A_G \rightarrow \mathbb{R}$. Denote by \( G_a \) the sub-digraph of \( G \) consisting of those arrows whose weights are less than \( a \).
 Naturally, the filtration of chain complexes $$\Omega(G_{a_1})\subseteq\Omega(G_{a_2})\subseteq \cdots\subseteq \Omega(G_{a_m}) = \Omega(G)$$ are induced by inclusion homomorphisms between path complexes
 $$\mathcal{P}(G_{a_1})\subseteq \mathcal{P}(G_{a_2})\subseteq \cdots\subseteq \mathcal{P}(G_{a_m}) = \mathcal{P}(G),$$ where \( \mathcal{P}(G_a) \) denotes the path complex consisting of all allowed paths in the sub-digraph \( G_a \) of \( G \).
\end{ex}
\begin{ex}[Path filtration]
    Let $A(G)$ be the path complex consisting of all allowed paths of $G$. Denoted $A^k(G)$ be the sub-path complex consists of paths of length less than or equal to $k.$ Then we obtain a series of increasing path complexes 
    $$\emptyset = A^{-1}(G)\subseteq A^0(G)\subseteq\cdots\subseteq A^m(G).$$
\end{ex}
\begin{ex}[Angle filtration~\cite{chen2023path}] Place the digraph $G$ into Euclidean space 
 $\mathbb{R}^3$ and map it onto unit 2-sphere $S^2=\{(\frac{2\pi t}{k},\frac{\pi s}{m})|0\leq t\leq k, 0\leq s\leq m\}$. Define an order $\leq$ on $S^2,$  $(t,s)\leq (t',s')$ if and only if $t < t'$ or, $t=t'$ and $s\leq s'$. Then there is an increasing digraph sequence
 $$\emptyset = G_{0,0}\subseteq G_{t_0,s_0}\subseteq \cdots\subseteq G_{k,m} =G,$$ which induces a filtration on path complexes $$\mathcal{P}(G_{0,0})\subseteq \mathcal{P}(G_{t_0,s_0})\subseteq \cdots\subseteq  \mathcal{P}(G_{k,m}) =\mathcal{P}(G),$$ where $G_{t,s}$ is the induced sub-digraph of $G$ spanned by $V_{G_{t,s}} = \{x \in V_G|~\frac{x}{||x||}\leq (t,s)\}$ and \( \mathcal{P}(G_{t,s}) \) denotes the path complex consisting of all allowed paths in \( G_{t,s} \).
\end{ex}
At this point, fix a filtration \( f: \mathcal{P} \rightarrow \mathbb{R} \) on the path complex \( \mathcal{P} \), the \emph{\( p \)-th persistent homology group} of \( \mathcal{P} \) from \( a \) to \( b \)~($a\leq b$) is defined as the image of the homomorphism  
\[
\varphi^{a,b}_{p} : H_p(\mathcal{P}_a) \rightarrow H_p(\mathcal{P}_b),
\]  
where \( \varphi^{a,b}_{p} \) is induced by the inclusion map \( i^{a,b} \colon \mathcal{P}_a \hookrightarrow \mathcal{P}_b \) between path complexes, with  
\[
\mathcal{P}_a = f^{-1}((-\infty, a]) \quad \text{and} \quad \mathcal{P}_b = f^{-1}((-\infty, b]).
\]  
That is, \( H^{a,b}_p = \mathrm{Im} \, \varphi^{a,b}_{p} \). The \emph{\( p \)-th persistent Betti number} from \( a \) to \( b \) is defined as the rank of \( H^{a,b}_p \), and is denoted by \( \beta^{a,b}_p \).

Furthermore, the persistent GLMY homology exhibit stability under noise permutations and its algorithm is studied in~\cite{chowdhury2018persistent,dey2022efficient}. This marks persistent GLMY homology as a practiable method to solve the real problems. Currently, persistent GLMY homology has been used to measure shape features and extract patterns from data that persist across multiple scales, with applications spanning biology, chemistry, complex networks, and complex diseases~\cite{che2024idopnetwork,chen2023persistent,chen2023path,chowdhury2019path,chowdhury2018functorial,wu2023metabolomic}. 

Additionally, the weighted persistent GLMY homology on digraphs and the (weighted) persistent homology for hypergraphs attracts increasing interests from protein-ligand binding affinity prediction, drug design and communicative network~\cite{chen2023persistent,gao2023persistent,liu2021hypergraph}. And now, it is still continuously delving into other advanced fields and disciplines.

\section{IdopNetwork meets GLMY theory}
\subsection{Approaches for analyzing network topology}

Network-based methods can systematically characterize the connections between different entities. Currently, many graph theory methods have been used to analyze the structure of complex networks~\cite{newman2003structure,west2001introduction}. These classic methods define network characteristics, such as degree centrality, betweenness centrality, small-worldness, and network homogeneity, to measure the similarity or difference of networks~\cite{rosvall2007information,wen2023regulating,xiao2021deciphering}. However, these measurements are usually limited to locality, based on the differences or correlations between nodes or edges, and rarely consider the overall topological structure of the network, such as connected components or holes~\cite{aktas2019persistence}. The basic idea of persistent homology is to replace data points by finding a set of parameterized simplicial complexes, which are usually composed of a union of points, edges, triangles, tetrahedrons, and high-dimensional polyhedrons, which encode changes in topological features~\cite{aktas2019persistence,ghrist2008barcodes}. Path homology, now renamed GLMY homology theory, has become a promising tool for analyzing and studying complex networks. GLMY homology theory is applied to weighted directed networks to capture directional information~\cite{lin2019weighted}. In this section, we introduce and list the current common network structure analysis methods and compare them with GLMY homology calculations to highlight the advantages of this theory (Figure 6).
\begin{figure}[htbp]
    \centering
\includegraphics[width=1\linewidth]{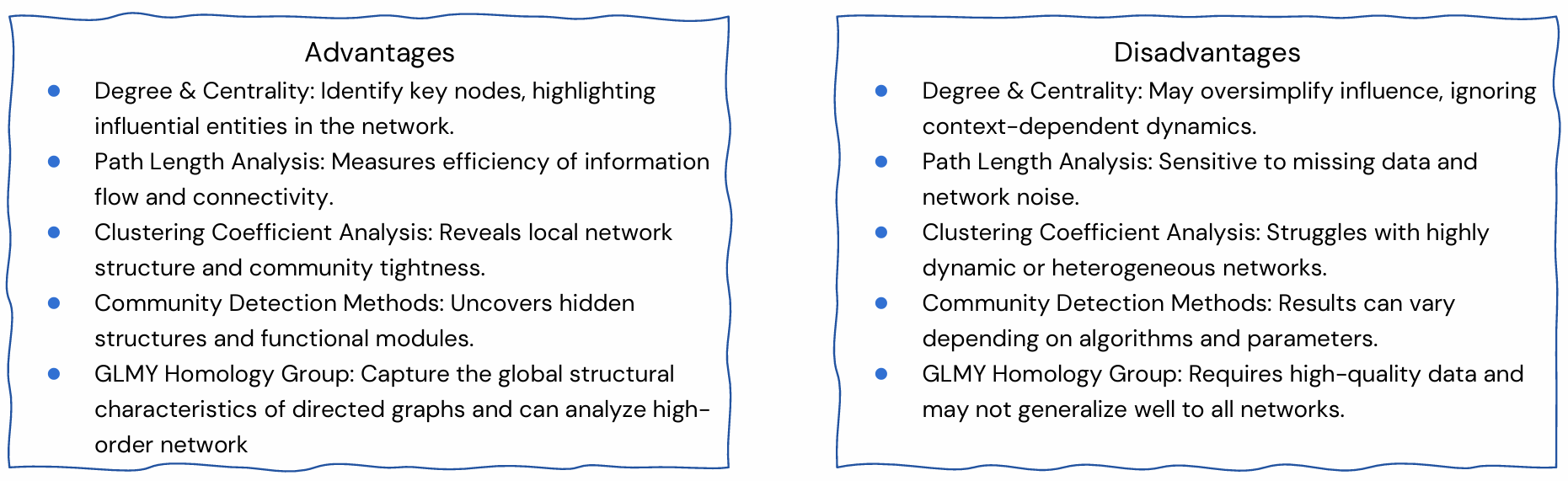}
    \caption{Advantages and disadvantages of different approaches for analyzing network topology}
    \label{fig:adanddisad}
\end{figure}
\subsubsection{Degree}
The degree of a node represents the number of edges directly connected to it, including metrics such as out-degree, in-degree, and average degree. Out-degree refers to the number of edges directed from a node to others, while in-degree represents edges directed toward the node from others. Average degree is the mean degree across all nodes in the network. In a weighted network, node degree can be expressed as out-strength and in-strength, accounting for edge weights~\cite{lindquist2011effective}. The degree distribution of the node is an indicator that describes the structure of the network, including Gaussian distribution, power-law distribution, etc. The power-law distribution shows that there are some highly connected nodes in the network, which is important for understanding the structure and properties of the network. 

The degree method is suitable for basic network analysis due to its simple and intuitive characteristics, but in the study of complex networks, its limitations may lead to a one-sided understanding of the importance of nodes. Node degree analysis is difficult to capture the global structural information of the network, and it also shows certain limitations when dealing with sparse networks.

\subsubsection{Centrality}
Centrality measures include degree centrality, betweenness centrality, and closeness centrality, which are used to measure the importance of nodes in a network~\cite{rodrigues2019network}. These indicators help identify key nodes in a network. Degree centrality is a standardized form of node degree, calculated by dividing the degree of a node by the maximum degree of nodes in the network. It is the most direct measure of node centrality in network analysis~\cite{rahim2018vehicular}. The larger the node degree of a node means that the node has more direct connections in the network, that is, the higher the degree centrality, the more important the node is in the network. Betweenness centrality node refers to the number of shortest paths through a node in a network~\cite{brandes2001faster}, which is a measure of the frequency of a node in all shortest paths in the network. Nodes with higher betweenness centrality are usually located in key positions in the network and are important for information transmission between different parts. Closeness centrality measures the average distance from a node to other nodes in the network. Nodes with higher closeness centrality are usually able to communicate with other nodes in the network faster~\cite{zhang2017degree}. 

Network centrality indicators play an important role in evaluating the importance of nodes, but they often ignore the dynamics of the network and have high computational complexity. A single centrality indicator is difficult to fully and accurately reveal the most critical information in the network, and a comprehensive analysis combining multiple indicators is required.
\subsubsection{Path length analysis }
Average shortest path length is a concept in network topology, defined as the average number of steps along the shortest path for all possible pairs of network nodes. It is a measure of the efficiency of information or mass transmission on a network. In large networks, many nodes are usually not connected to each other, which may make the distance between two nodes be regarded as infinite, resulting in the average path length of the entire network being infinite. Mao and Zhang~\cite{mao2013analysis} proposed a network topology description method that can define the average path length of the network as the average distance between a pair of interconnected nodes, which eliminates the data divergence problem generated during the calculation process. 

Path length analysis provides valuable global structural information in many network analyses, but it may ignore the importance of nodes when dealing with dynamic, weighted, sparse or complex networks, and is sensitive to network noise. Therefore, in practical applications, path length analysis usually needs to be combined with other network analysis methods.
\subsubsection{Clustering coefficient analysis}
The clustering coefficient of the network topology is an indicator that measures the degree of node aggregation in the network. It describes the closeness of the interconnection between neighboring nodes in the network. Clustering coefficient analysis is important for understanding social structure, modular organization, and information dissemination in networks~\cite{masuda2018clustering}. 

 Kartun-Giles and Bianconi~\cite{kartun2019beyond} performed a large-scale statistical analysis of the node neighborhood topology of real networks by constructing clique complexes of the network and calculating Betti numbers. The results show significant differences between the node neighborhood topology of a real network and the stochastic topology of a stochastic simplicial complex null model, thus revealing the local organizing principles of node neighborhoods. Large-scale statistical analysis of topological properties of node neighborhoods enables a clear distinction between power-law networks and planar road networks~\cite{kartun2019beyond}. 
 
Clustering coefficient analysis is important in the study of local structure of networks, but it lacks the ability to identify cross-level clusters, cannot process weighted data, and cannot reflect the true structural characteristics of the network.

\subsubsection{Community detection methods }
Community detection methods, namely network modularity, reveal the clustered organizational structure and functional modules of nodes in the network by identifying subgraphs or communities in the network. Lancichinetti and Fortunato~\cite{lancichinetti2009community} tested several methods with heterogeneous distribution of community sizes on benchmark graphs. These methods were tested on Newman's benchmark and random graphs~\cite{newman2006modularity}. The results showed that the method has excellent performance and low complexity in calculation. This advantage is helpful for analyzing large network systems~\cite{lancichinetti2009community}. 

Community detection methods play a role in revealing network structure and function, but their algorithms have high requirements on network density and initial calculation values, and usually can only reveal the local optimal structure of the network, but are difficult to explain from a global perspective.

\subsubsection{GLMY homology group}
The GLMY homology group is an essensial and powerful algebraic invariant in algebraic topology that reflects and reveals the global structural characteristics of digraphs~\cite{grigor2012homologies} with  computational properties~\cite{chowdhury2018persistent}. The concept was initially introduced by Grigor'yan et al. in 2012 and was subsequently renamed GLMY homology group. Due to the broad applications of GLMY homology theory in biology, chemistry, complex networks and complex diseases, as well as its theoretical breakthroughs in terms of discretization of topological spaces, an increasing number of researchers from both theoretical and applied fields, including statisticians, have devoted themselves to study and employ the GLMY theory, making it an emerging interdisciplinary field over just the past decade.

Compared to classical homology theory, GLMY homology groups have broader applicability, as most complex systems can be represented by graphs or digraphs. Additionally, GLMY homology group can more precisely track and simulate the asymmetric and unbalanced relationships between objects in complex systems by taking direction and strength of digraphs into account. Finally, by analyzing propagation pathways, the theory makes the in-depth exploration of the high-dimensional topology and geometric structure of systems possible. It helps identify key connections and primary conflicts within the network, thereby facilitating the precise discovery of the relationship between structure and function in networks. It even provides more personalized interpretation and translation for each complex network. 

In summary, GLMY homology group, as a method of attaching algebraic invariants to digraphs, has higher adaptability and breadth than traditional network topology analysis methods due to its strong theoretical foundation and broad application prospects. It provides important theoretical support in multi-dimensional network analysis, network structural understanding and function mining, and especially shows unique advantages in the dynamic evolution analysis of complex networks.

\subsection{Applications in IdopNetwork with GLMY}
The network can represent the interaction relationship between genes from the system level and extract and analyze the rich information it contains from the topological structure~\cite{newman2003structure,west2001introduction}. GLMY homology, as a method to explore the topological structure of complex paths, can be effectively applied to the network structure analysis based on directed graphs, and the key information hidden in the complex network can be mined by comparing the topological features in the different network. 

At present, idopNetwork has been widely used in biological and medical research. In 2020, Sun et al.~\cite{cancers12082086} used the model to calculate the interaction between neuroblastoma genes. The results of the network characterized the biological processes of gene interactions in response to developmental and environmental changes, and finally located biomarkers associated with neuroblastoma risk. Cao et al.~\cite{cao2022modeling} used the model to map the spatial relationship between intestinal microbiota and ulcerative colitis and clarified the mechanism by which the spatial variation of microorganisms determines the impact of intestinal microbiota on human health. Subsequently, Wang et al.~\cite{wang2022vaginal} mined the characteristics of the mechanism predictors of aerobic vaginitis through idopNetwork to characterize the causal interdependence between microbial interactions and AV and determined the topological changes of the vaginal microbiota network. Dong et al.~\cite{dong2021genomic} used the model as a tool to reveal the pathogenesis and transmission of COVID-19, develop effective containment and mitigation strategies, and screen and classify more susceptible populations and asymptomatic carriers. 

In 2023, Wu et al.~\cite{wu2023metabolomic} introduced the GLMY homology theory into idopNetwork for the first time to analyze the topological characteristics of network results, thereby revealing the information flow in the intestinal inflammatory metabolic network (Figure 7). Wu et al. broke through the traditional reductionism that focuses on the genetic mechanism analysis of a single biological molecule. From the perspective of systems biology, they regarded the biological molecules in the complex disease process. By applying the idopNetwork model to the multi-omics data analysis of complex diseases, Wu et al. analyzed the formation mechanism and pathogenesis of complex diseases, providing new research ideas for clinical treatment and drug design. In addition, they applied the GLMY homology theory to network topological structure analysis for the first time and compared the topological characteristics of different networks in different dimensions. This analysis method can more accurately compare the differences between networks, provide innovative theoretical support for subsequent network result analysis, and promote the further development of complex network structure research. 
\begin{figure}[htbp]
    \centering
\includegraphics[width=1\linewidth]{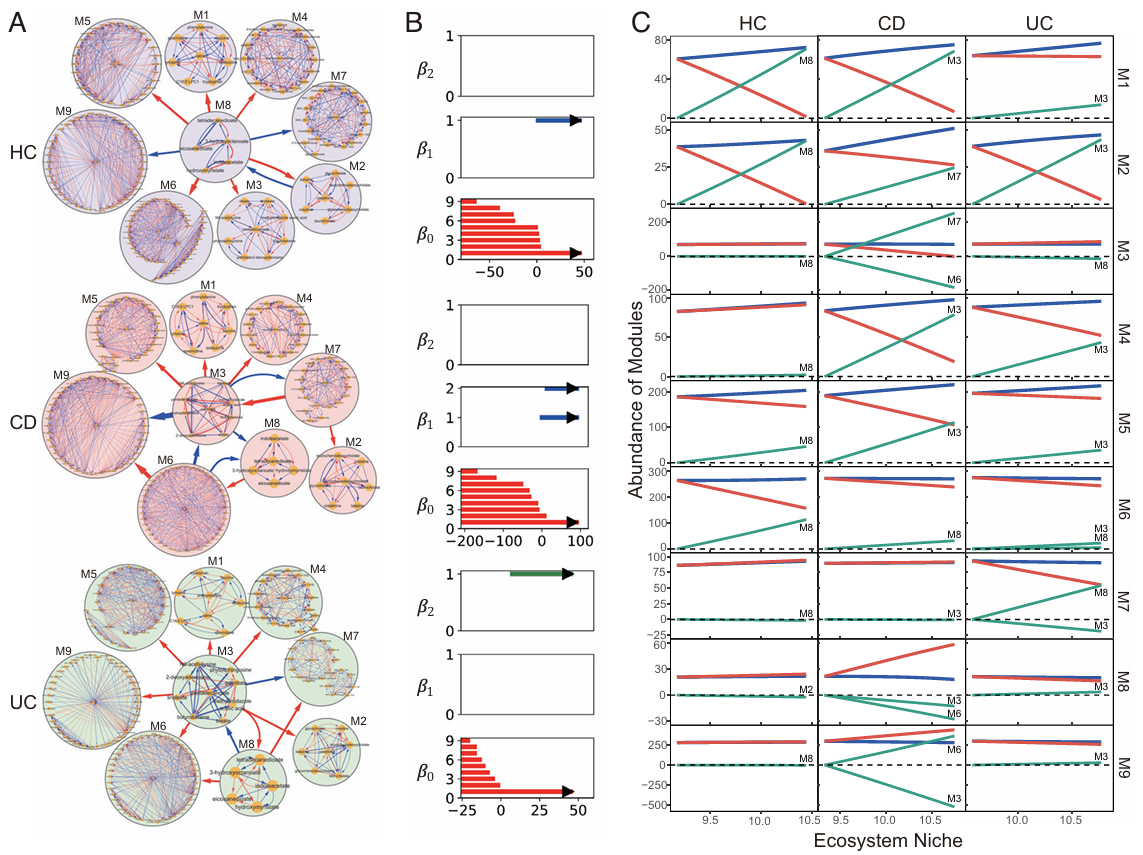}
    \caption{Application of idopNetwork meets GLMY. Reproduced from Wu et al., 2023.}
    \label{fig:wu2023}
\end{figure}

Subsequently, the combination of idopNetwork and GLMY homology theory has attracted widespread attention, leading to extensive research by scholars. Che et al.~\cite{che2024idopnetwork} applied this combination in drug design research, uncovering how genes and proteins mediate crosstalk between cells to shape the body’s response to drugs through the combination of idopNetwork and GLMY theory (Figure 8). Meanwhile, Gong et al.~\cite{gong2024topological} used this theory to explore the impact of thinning on forestry soil microorganisms (Figure 9). By integrating GLMY homology theory, they analyzed the topological structure of the omnidirectional network and identified key microbial interaction pathways that regulate the structure and function of soil microbial communities.
\begin{figure}[htbp]
    \centering
\includegraphics[width=1\linewidth]{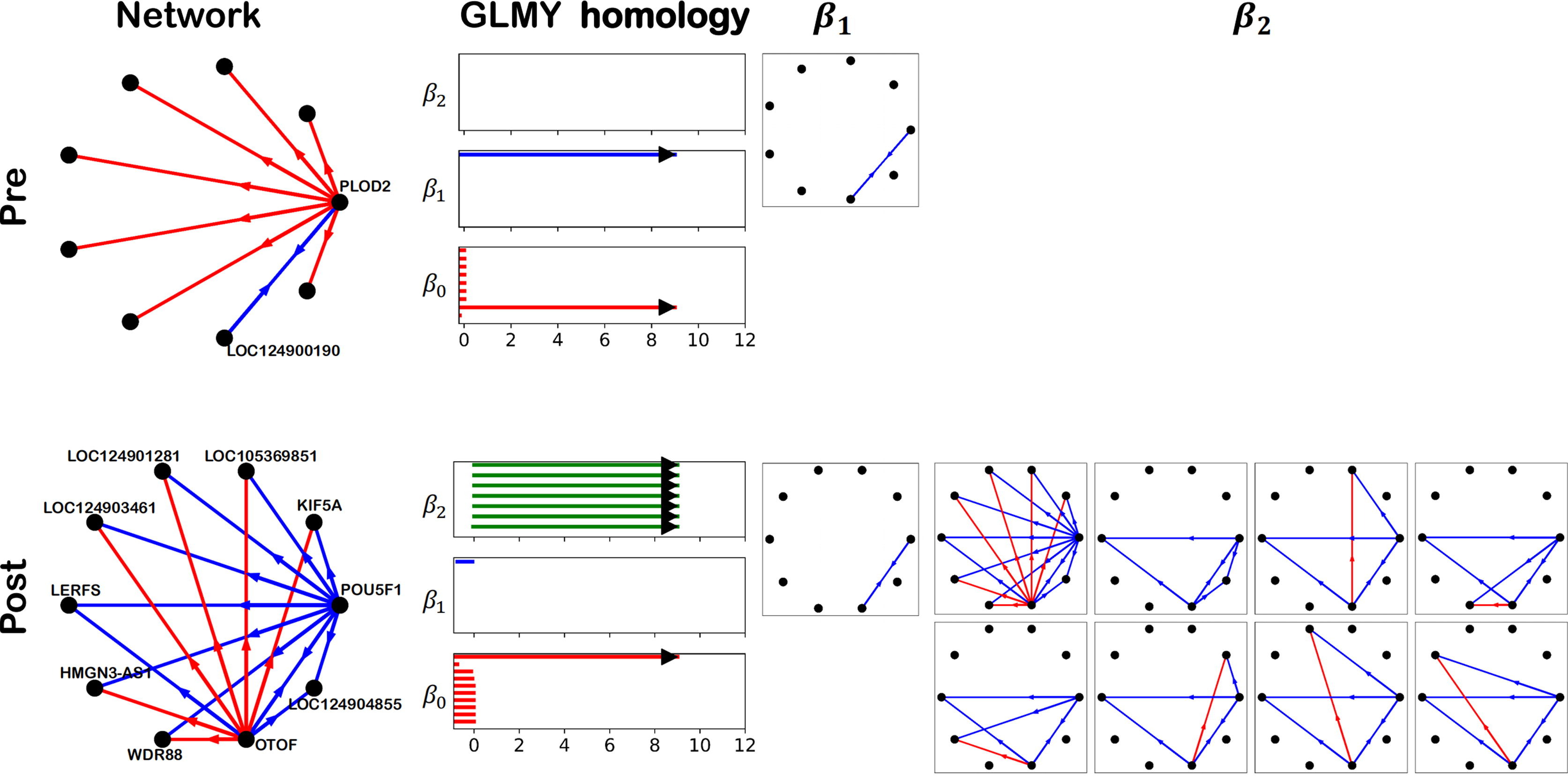}
    \caption{Application of idopNetwork meets GLMY. Reproduced from Che et al., 2024.}
    \label{fig:che2024}
\end{figure}
\begin{figure}[htbp]
    \centering
\includegraphics[width=1\linewidth]{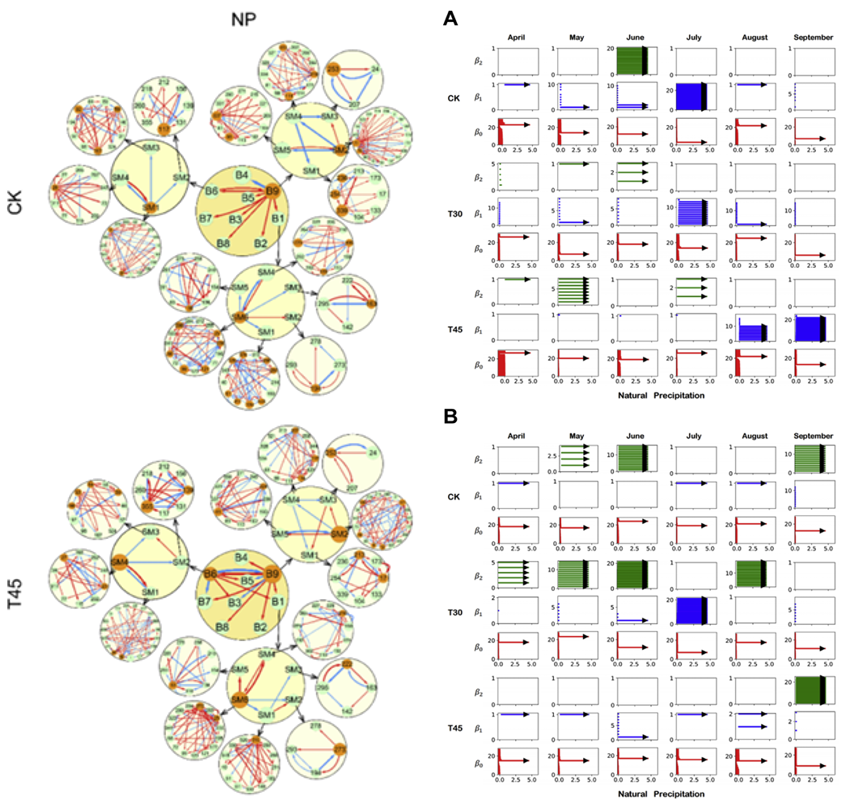}
    \caption{Application of idopNetwork meets GLMY. Reproduced from Gong et al., 2024.}
    \label{fig:gong2024}
\end{figure}

\section{Conclusions and outlook}
A large amount of evidence shows that the heterogeneity inside and outside the system gives the system the ability to adapt to changes in the internal and external environment through specific regulatory mechanisms. This adaptability is driven by the dynamic interactions between different entities in the system, including mutually beneficial cooperation and competitive selection among entities~\cite{vitale2021intratumoral}. The interactions between entities and their spatiotemporal distribution in the system and the environment are crucial to the functional performance of the system under environmental changes or external interventions. In order to deeply understand this complex interaction, network science and complex system analysis tools provide strong support. 

Although existing algorithmic tools have made significant strides in analyzing interactions between different entities and characterizing the complex network structures within systems, these methods are difficult to fully capture the intensity, causality and symbols of interactions between entities. To address these challenges, researchers introduced the idopNetwork model and combined it with the GLMY homology method to conduct an in-depth analysis of network topology. The GLMY homology, as a network topology analysis tool, reveals higher-order structural characteristics and hidden information flows within complex networks. And it deserves to say that this model captures the nonlinear characteristics of interactions, it remains limited by its static representation of the network and has difficulty capturing dynamic topological features when calculating continuous interactions.

By combining idopNetwork with GLMY homology, the researchers overcame the limitations of traditional methods and achieved a more comprehensive analysis of the interactions and information transfer between entities. This progress not only provides a new perspective for reconstructing interaction networks in complex systems but also opens a new research direction for precise control and optimization of complex system functions and is expected to promote the realization of precise control and complex system modeling in more application scenarios. 

Currently, the combination of GLMY homology and idopNetwork has demonstrated its effective capabilities in modeling, probing, tracking and predicting the structure and function of complex networks. Inspired by this, it is reasonable to believe that combining the idopNetwork with weighted or multi-edge digraphs and hypergraphs will provide us with more sophisticated and effective tools and methods to address a wider range of scientific bottlenecks and challenges. 

\section*{Acknowledgements}
This work is partially supported by the start-up research funds of authors from Beijing Institute of Mathematical Sciences and Applications~(BIMSA). The authors are also very grateful to BIMSA and anonymous referees for many helpful comments.

%\bibliographystyle{plain}        % 或 alpha, abbrv, etc.
%\bibliography{references}        % references.bib 文件名

\end{document}